\documentclass[11pt]{article}
\usepackage{amssymb,amsmath,cite,geometry,graphicx,url}
\catcode`\@=11

\@addtoreset{equation}{section}
\def\theequation{\arabic{section}.\arabic{equation}}
     
\catcode`\@=11
\def\thesection{\arabic{section}.}

\def\appendix{\setcounter{section}{0}
        \def\thesection{Appendix.}
        \def\theequation{\Alph{section}.\arabic{equation}}}
\def\section{\@startsection{section}{1}{\z@}{3.5ex plus 1ex minus
   .2ex}{2.3ex plus .2ex}{\large\bf}}
     
\long\def\@makefntext#1{\parindent 0cm\noindent
\hbox to 1em{\hss$^{\@thefnmark}$}#1}

\newcommand{\captionfonts}{\small}
\makeatletter  
\long\def\@makecaption#1#2{%
  \vskip\abovecaptionskip
  \sbox\@tempboxa{{\captionfonts #1: #2}}%
  \ifdim \wd\@tempboxa >\hsize
    {\captionfonts #1: #2\par}
  \else
    \hbox to\hsize{\hfil\box\@tempboxa\hfil}%
  \fi
  \vskip\belowcaptionskip}
\makeatother   

\begin{document}
\begin{titlepage}
\vspace{.5in}
\begin{flushright}
March 2022; updated May 2022\\  
\end{flushright}
\vspace{.5in}
\begin{center}
{\Large\bf
 A Schwarzian on the Stretched Horizon}\\  
\vspace{.4in}
{S.~C{\sc arlip}\footnote{\it email: carlip@physics.ucdavis.edu\\\hspace*{.85em}ORCID iD 0000-0002-9666-384X}\\
        {\small\it Department of Physics}\\
       {\small\it University of California}\\
       {\small\it Davis, CA 95616}\\{\small\it USA}}
\end{center}

\vspace{.5in}
\begin{center}
{\large\bf Abstract}
\end{center}
\begin{center}
\begin{minipage}{4.75in}
{\small
It is well known that the Euclidean black hole action has a boundary term at the horizon
proportional to the area.  I show that if the horizon is replaced by a stretched horizon
with appropriate boundary conditions, a new boundary term appears, described 
by a Schwarzian action similar to the recently discovered boundary actions in ``nearly 
anti-de Sitter'' gravity.
}
\end{minipage}
\end{center}
\end{titlepage}
\addtocounter{footnote}{-1}

\section{Introduction}

It has long be recognized that ``boundary'' degrees of freedom, either at infinity or at or near the 
horizon, may play a central role in black hole thermodynamics  \cite{York,Suss,Strominger,Pada,Padb}.  
In the path integral, the entropy of the ``Euclidean'' black hole---the stationary black hole analytically 
continued to Riemannian signature---comes from a boundary term, either at infinity \cite{Gibbons} or
 at the horizon \cite{Teitelboim1,HawkingHorowitz}; in the latter case, it is
 canonically conjugate to a horizon deficit angle \cite{CarTeit}.  At the
microscopic level, the symmetries of the horizon suggest the existence of boundary degrees of
freedom as ``would-be diffeomorphisms,'' deformations that would ordinarily be pure gauge but 
become physical because boundary conditions restrict the true gauge transformation \cite{Cara,Carx,Padc}.  
In some cases, the argument for such degrees of freedom is quite strong.  The (2+1)-dimensional BTZ 
black hole, for instance, has no local bulk degrees of freedom, but one can still explicitly construct an 
induced dynamical action at the boundary that may account for the Bekenstein-Hawking entropy 
\cite{Coussaert,BTZd}.  

Recently, related boundary actions have proven important in a somewhat different
context.  In two-dimensional ``nearly anti-de Sitter space''  \cite{AlPol,Mal,Bloem}, the boundary of
asymptotically AdS space at infinity is replaced  by a finite boundary.  A bulk action given by
Jackiw-Teitelboim gravity \cite{Jackiw,Teit} or its variants then induces a boundary action that
can be described by a Schwarzian,
\begin{align}
I = C\int \!d\tau\,\phi \{f,\tau\}  \, ,
\label{a1}
\end{align}
where the Schwarzian derivative $\{f,\tau\}$ is
\begin{align}
\{f,\tau\} =  \frac{\dddot{f}}{\dot{f}} - \frac{3}{2}\frac{\ddot{f}^2}{\dot{f}^2}
\label{a2}
\end{align}
(a dot is a derivative with respect to $\tau$).  This is a powerful result: the Schwarzian action can be quantized
 exactly \cite{Verlinde,Mertens}, and has fascinating connections to a variety of conformal field theories 
and matrix models \cite{SYK,SYKb,Saad}.  

While most of the recent work on the Schwarzian action has taken place in the context of nearly 
anti-de Sitter space, the same action also appears elsewhere:  in ``nearly de Sitter space'' \cite{Malb}, for instance,
and as a corner term in asymptotically flat (2+1)-dimensional gravity \cite{Carlipx}.    The Schwarzian
action is closely related to Liouville theory \cite{Mertens}, which is ubiquitous in quantum gravity,
and there are arguments from effective field theory that a Schwarzian at a one-dimensional boundary is 
generic \cite{Larsen}.  So it may not be surprising to find similar actions elsewhere.

In particular, nearly anti-de Sitter space approximates the near horizon geometry 
of an extremal black hole, and it seems natural to ask whether there is an extension to more generic  
black holes.  In this paper, I will show that a Schwarzian action does, in fact, describe the stretched 
horizon of an arbitrary nonextremal Euclidean black hole, albeit with slightly different boundary conditions.
This opens up the possibility that some of the powerful results from nearly anti-de Sitter space may be
applicable to this broader setting.

\section{The boundary action}

To understand the appearance of the Schwarzian action, we will first need two ingredients, the dimensional
reduction of the near-horizon region and the form of the near-horizon metric.  Both are fairly standard:
\begin{itemize}
\item {\bf Near-horizon dimensional reduction}\\[1ex]
It has been understood for some time that the near-horizon region of an extremal or nearly extremal black
hole can be described by a two-dimensional dilaton gravity model \cite{KerrCFT}.  While it is not quite as
well known, the same is true for the near-horizon region of an arbitrary stationary black hole \cite{Yoon,Carb}.
More precisely, write the metric near the horizon in the general form
\begin{align}
ds^2 = g_{ab}dx^adx^b + \phi_{\mu\nu}(dy^\mu + A_a{}^\mu dx^a)(dy^\nu + A_b{}^\nu dx^b) \, ,
\label{b1}
\end{align}
where lower case Roman indices (a,b,\dots) run from $0$ to $1$ and label the ``$r$--$t$ plane,'' while
lower case Greek indices ($\mu$,$\nu$,\dots) run from $2$ to $D-1$ and label the transverse coordinates.  
Then as shown in \cite{Carb}, after a conformal rescaling of the metric, the near-horizon behavior is described 
by a two-dimensional action
\begin{align}
I_2 = \frac{1}{16\pi G}\int_M\!d^2x\,\sqrt{g}\left\{ \varphi R  + V[\varphi]\right\} + \dots \, ,
\label{b2}
\end{align}    
where 
$$\varphi = \sqrt{|\det\phi_{\mu\nu}|}$$
is the transverse volume element.  The omitted terms are additional Kaluza-Klein-like matter fields, which couple 
to the the two-dimensional metric and dilaton.  But as shown in \cite{Carb}, these vanish at the horizon and are very 
small in the near-horizon region, making them irrelevant for the phenomena considered here.   

As always, if the metric is fixed at a boundary, an extra Gibbons-Hawking boundary term is also
required \cite{Gibbons}.  Its dimensionally reduced form is
\begin{align}
I_{bdry} = \frac{1}{8\pi G}\int_{\partial M}\!dx\sqrt{h}\,\varphi K   \, ,
\label{b3}
\end{align}
where $h$ is the induced metric on the boundary $\partial M$ and $K$ is the trace of the extrinsic curvature
of $\partial M$.

\item {\bf The near-horizon metric}\\[1ex]
We will also need the near-horizon form of the metric, analytically continued to Riemannian signature.
For a Schwarzschild black hole, it is well known that the dimensionally reduced metric near the horizon takes
the form
\begin{align}
 ds^2 = -N^2dt^2 + N^{-2}dr^2 \quad \hbox{with $N=\sqrt{2\kappa(r-r_+)}$} \, ,
 \label{b4}
 \end{align}
 where $\kappa$ is the surface gravity and the horizon is located at $r=r_+$.  This  turns out to be generic: the same 
 form occurs  for an arbitrary stationary nonextremal black hole, regardless of the presence of matter or even the
 detailed form of the field equations \cite{Visser,Pada}.  If one now sets
 \begin{align}
 r = r_+ + \frac{1}{2}\kappa\rho^2, \ \tau = i\kappa t  \, ,
 \label{b5}
 \end{align}
 the near-horizon metric becomes\footnote{Technically, the two-dimensional metric in (\ref{b2}) is conformally
 rescaled from the $D$-dimensional expression, but near the horizon the effect is higher order in $\rho$ \cite{Carb}.}
 \begin{align}
 ds^2 = d\rho^2 + \rho^2d\tau^2 \, .
 \label{b6}
 \end{align}
The horizon is at $\rho=0$, and the coordinate $\rho$ now has a clear physical meaning as the proper distance 
from the horizon, while the ``Euclidean time''  $\tau$ is periodic with a period of $2\pi - \Theta$, where the deficit 
angle $\Theta$ is conjugate to the transverse area $\varphi$ and vanishes on shell \cite{CarTeit}.
\end{itemize}
 
We can now compute the boundary action (\ref{b3}) for a stretched horizon.  We enlarge the horizon by considering 
a small closed path $\Delta$ encircling the origin.  Such a path can be described as a parametrized curve 
$(\rho(\sigma),\tau(\sigma))$, where for later convenience we will take $\sigma$ to be periodic with a fixed 
range $[0,1]$.  The length of $\Delta$ is then
\begin{align}
\ell = \oint\!ds = \int_0^1\!\varepsilon \,d\sigma \quad \hbox{with}\quad
\varepsilon = \frac{ds}{d\sigma} = \left(\rho^{\prime 2} + \rho^2\tau^{\prime 2}\right)^{1/2} \, ,
\label{b7}
\end{align}
where a prime denotes a derivative with respect to $\sigma$.  For $\ell$ to be small, $\varepsilon$ must 
be small, which in turn implies a small $\rho$; that is, $\Delta$ must remain close to the horizon.

Note that by the smooth Schoenflies theorem, a two-dimensional diffeomorphism can map any such curve into 
any other.  But we will treat $\Delta$ as a boundary, and the gravitational action is not invariant under diffeomorphisms
transverse to a boundary.  In this sense, different choices of the path $\Delta$ represent the ``would-be
diffeomorphisms'' of \cite{Cara}.

The unit tangent and normal vectors to $\Delta$ are
\begin{align}
t^a & = \frac{1}{\varepsilon}\left(\rho',\tau'\right)  \nonumber\\
n^a & = \frac{\rho}{\varepsilon}\left(\tau',-\frac{\rho'}{\rho^2}\right)  \, .
\label{b8}
\end{align}
The extrinsic curvature is thus
\begin{align}
K = t^at^b\nabla_an_b 
   = \frac{1}{\varepsilon^3}\left(\rho\rho^\prime\tau^{\prime\prime} - \rho\tau^\prime\rho^{\prime\prime}
   + 2\rho^{\prime 2}\tau^\prime + \rho^2\tau^{\prime 3}\right)  \, .
\label{b9}
\end{align}
The induced metric on $\Delta$ is $\sqrt{h} = \varepsilon$, and with a little calculation, one can show that
\begin{align}
\sqrt{h}K = \tau' 
   - \frac{1}{\sqrt{\varepsilon^2-\rho^{\prime 2}}}\left(\rho^{\prime\prime} - \rho'\frac{\varepsilon'}{\varepsilon}\right)
   = \tau' - \left(1 - \frac{\rho^{\prime 2}}{\varepsilon^2}\right)^{-1/2}\left(\frac{\rho'}{\varepsilon}\right)^\prime \, .
\label{b10}
\end{align}
Hence the boundary action (\ref{b3}) takes the form
\begin{align}
I_{bdry} = \frac{1}{8\pi G}\int_\Delta \!d\sigma\, \varphi\left(\tau' 
  - \left(1 - \frac{\rho^{\prime 2}}{\varepsilon^2}\right)^{-1/2}\left(\frac{\rho'}{\varepsilon}\right)^\prime\right) \, .
\label{b11}
\end{align}

The second term in parentheses is a total derivative.  If we choose $\varphi$ to be constant on $\Delta$, 
we therefore have
\begin{align}
I_{bdry} 
  = \frac{1}{8\pi G}\int_\Delta \!d\sigma\, \varphi \tau' = \frac{\varphi|_\Delta}{4G}\left(1-\frac{\Theta}{2\pi}\right) \, ,
\label{b12}
\end{align}
which is the standard boundary action; the only effect of enlarging the horizon is to slightly shift the value of
the area $\varphi|_\Delta$.   This is, of course, an approximation---the metric (\ref{b6}) has corrections away from 
the horizon---but any such corrections will be of order $\rho^2$.  

Now, observe that $\varepsilon$ appears in (\ref{b11}) only in the combination $\varepsilon d\sigma = ds$.  By 
a suitable choice of parametrization, we can take $\varepsilon$ to be constant on $\Delta$.  In that case, from 
(\ref{b7}), $\varepsilon$ is just the length of the stretched horizon, while $\sigma$ is a rescaled proper length.

\section{A Schwarzian at the horizon}

Let us now drop the requirement that $\varphi$ be constant on $\Delta$.  We will, however, assume that $\Delta$ 
is ``not too irregular.''   More precisely, while $\rho$ is $\mathcal{O}(\varepsilon)$, we will now assume that 
$\rho'$ is $\mathcal{O}(\varepsilon^2)$.  Then from (\ref{b7}),
\begin{align}
\frac{\rho}{\varepsilon} = \frac{1}{\tau'}\left(1 - \frac{\rho^{\prime 2}}{\varepsilon^2}\right)^{1/2}
\approx \frac{1}{\tau'}\left(1 - \frac{1}{2}\frac{\rho^{\prime 2}}{\varepsilon^2}\right)
\approx  \frac{1}{\tau'}\left(1 - \frac{1}{2}\frac{\tau^{\prime\prime\,2}}{\tau^{\prime\,4}}\right) \, .
\label{c1}
\end{align}
To lowest order, with $\varepsilon$ held fixed, the boundary action (\ref{b11}) then becomes
\begin{align}
I_{bdry} = \frac{1}{8\pi G}\int_\Delta \!d\sigma\, 
   \varphi\left(\tau' + \left(\frac{\tau^{\prime\prime}}{\tau^{\prime 2}}\right)^\prime \right) 
   = \frac{1}{8\pi G}\int_\Delta \!d\tau \, 
   \varphi\left(1 - \frac{1}{2}\frac{\tau^{\prime\prime\,2}}{\tau^{\prime\,4}} +\frac{1}{\tau^{\prime\,2}}\{\tau,\sigma\}\right) \, .
\label{c2}
\end{align}
(Note that this approximation only requires the lowest order term in (\ref{c1}); the next order term will be important
below.)

We evidently have a problem: the piece we are interested in, the Schwarzian derivative, is a small correction to 
the standard area term (\ref{b12}), and can be modified by making small changes to the leading term.   In 
particular,  the Schwarzian derivative transforms anomalously: under a reparametrization
 $\sigma\rightarrow {\bar\sigma}(\sigma)$, 
\begin{align}
\left(\frac{d\tau}{d\sigma}\right)^{-2}\{\tau,\sigma\} = \left(\frac{d\tau}{d{\bar\sigma}}\right)^{-2}\{\tau,{\bar\sigma}\} 
   + \frac{1}{\tau^{\prime\,2}}\{{\bar\sigma},\sigma\}  \, .
\label{c2a}
\end{align}
Thus in the integrand in (\ref{c2}),
$$ 
1 - \frac{1}{2}\frac{\tau^{\prime\prime\,2}}{\tau^{\prime\,4}} \rightarrow
1 - \frac{1}{2}\frac{\tau^{\prime\prime\,2}}{\tau^{\prime\,4}} +  \frac{1}{\tau^{\prime\,2}}\{{\bar\sigma},\sigma\} \, ,
$$
and we can certainly find a new parametrization for which this term reduces to $1$.  
Such a choice seems rather arbitrary, though.  The parameters we are using are physically natural: $\tau$ is 
the background Killing time, and $\sigma$ is the scaled proper length.  So the question remains whether 
the particular structure in (\ref{c2}) has any deeper meaning.

For the particular case of the near-horizon black hole, it does.  To see this, we will need two additional elements.
First, as noted above, if $\varphi$ is constant the boundary term reduces to the usual area factor, with 
no extra dynamics.  This suggests that we might split off a constant  part of $\varphi$ to 
separate out the leading contribution.   From (\ref{b5}), it is clear that the first nonconstant piece of $\varphi$ 
appears at order $\rho^2$, so to the order we are considering,
\begin{align}
\varphi = \varphi_+ +  \frac{1}{2}\rho\partial_\rho\varphi  \, .
\label{c3}
\end{align}
The boundary term (\ref{c2}) is thus
\begin{align}
I_{bdry} = \frac{1}{8\pi G}\int_\Delta \!d\tau\, \varphi_+
   + \frac{1}{16\pi G}\int_\Delta \!d\tau\, \rho\partial_\rho\varphi
   \left(1 - \frac{1}{2}\frac{\tau^{\prime\prime\,2}}{\tau^{\prime\,4}} + \frac{1}{\tau^{\prime\,2}}\{\tau,\sigma\}\right) \, .
\label{c4}
\end{align}

Second, let us reconsider the boundary conditions at $\Delta$.  The full boundary term in the variation 
of the action $I_2+I_{bdry}$ is \cite{Goel}
\begin{align}
\delta(I_2+I_{bdry}) = \hbox{\em equations of motion\ } + \frac{1}{8\pi G}\int_\Delta \!d\sigma\, 
    \left[ K\sqrt{h}\,\delta\varphi + n^a\partial_a\varphi\,\delta(\sqrt{h})\right] \, .
\label{c5}
\end{align}
(The same result can be obtained in the Lorentzian setting from the symplectic form in \cite{Carb},
and is a special case of the results of \cite{Cadoni}.)
The boundary conditions we have assumed so far are the standard Dirichlet conditions, in which 
$\varphi$ and $\sqrt{h}$ are fixed.  This is certainly a reasonable choice, especially if one is 
taking the two-dimensional model to be fundamental.  

For an action obtained by dimensional reduction, though, it seems equally natural to fix the transverse variables 
$\varphi$ and  $n^a\nabla_a\varphi$.  Geometrically, this is a ``free boundary'' condition: the full geometry
of the stretched horizon, intrinsic (given by $h$) and extrinsic (given by $K$), is unrestricted, 
while the transverse area profile is fixed.  In the nearly anti-de Sitter case, the metric $h$ determines
the inverse temperature, and fixing its conjugate $\partial_n\varphi$ leads to a microcanonical 
ensemble with fixed ADM mass \cite{Goel}.  For the nonextremal case this is trickier, since the ADM 
mass is much more complicated \cite{Kunstatter}.  It is still true, however, that $\nabla_n\varphi$ 
remains conjugate to the length of the horizon in imaginary time, and thus conjugate to the inverse
 temperature $\beta$.

This new choice of boundary conditions requires an additional boundary term.  From (\ref{c5}),
\begin{align}
I^{(2)}_{bdry} = -\frac{1}{8\pi G}\int_\Delta \!d\sigma\, n^a\nabla_a\varphi  \sqrt{h}  
   = -\frac{1}{8\pi G}\int_\Delta \!d\sigma\, \left[%
   \rho\tau' \partial_\rho\varphi 
   - \frac{\varepsilon\rho'}{\rho^2}\partial_\tau\varphi\right]
   = -\frac{1}{8\pi G}\int_\Delta \!d\tau\, \rho\partial_\rho\varphi  \, ,
\label{c6}
\end{align}
where I have used (\ref{b8}) and the fact that $\partial_\tau\varphi=0$ for a stationary background.%
\footnote{The boundary value $\varphi|_\Delta$ can still depend on $\tau$, of course, but only through the 
dependence of $\Delta$ on $\tau$.}

Now combine the two boundary actions (\ref{c4}) and (\ref{c6}).   From  (\ref{c1}),  
to the order we are considering,
\begin{multline}
I_{bdry}  + I^{(2)}_{bdry} = \frac{1}{8\pi G}\int_\Delta \!d\tau\, \varphi_+
   - \frac{1}{16\pi G}\int_\Delta \!d\tau\, \rho\partial_\rho\varphi
   \left( 1 + \frac{1}{2}\frac{\tau^{\prime\prime\, 2}}{\tau^{\prime 4}} 
   - \frac{1}{\tau^{\prime\,2}}\{\tau,\sigma\}\right) \\[1ex]
   =  \frac{\varphi|_+}{4G}\left(1-\frac{\Theta}{2\pi}\right)
   - \frac{1}{16\pi G}\int_\Delta \!d\sigma\, \frac{\varepsilon}{\tau^{\prime\,2}}\partial_\rho\varphi
   \left( \tau^{\prime\,2} - \{\tau,\sigma\}\right) \, .
 \label{c7}
\end{multline}
The first term in (\ref{c7}) is the standard transverse area contribution.  The second is essentially a Schwarzian 
action, with a coefficient that depends only on $\partial_\rho\varphi$.  Thus by changing our boundary conditions 
and switching our background source from $\varphi$ to $\partial_\rho\varphi$, we have isolated a distinct, 
and interesting, second order piece of the action.

\section{Conclusion}

Obtaining a horizon action is, of course, only a first step.  The Schwarzian action has been extensively studied 
over the past few years in the nearly anti-de Sitter \cite{AlPol,Mal} and nearly de Sitter \cite{Malb} settings, and it 
seems plausible that some of those results can be adapted to this new case.  There may be subtleties, though:
the flat space metric (\ref{b6}) is not the metric anti-de Sitter space, and while the action (\ref{c7}) is almost the standard 
Schwarzian, it differs by an extra prefactor of $1/\tau^{\prime\,2}$.   One might learn more by investigating the extremal 
limit  of the action (\ref{c7}), where the expansion (\ref{c3}) breaks down; it should presumably be possible to
reproduce known results. 

It is tempting to try to extend this derivation to higher orders, to see whether one continues to obtain functions of the 
Schwarzian, as occurs in the nearly AdS case \cite{Iliesiu}.  This is tricky, though: the near-horizon metric (\ref{b6}) 
has $\mathcal{O}(\rho^4)$ corrections that would need to be accounted for, and universality would probably be lost.
It should be possible to repeat this derivation in Lorentzian signature, although the definition of a stretched horizon 
becomes a bit delicate there \cite{Carc}.  In \cite{Goel}, a collection of alternative boundary conditions is described;  
the corresponding actions will probably not be Schwarzians, but it could be worthwhile to understand the differences.  
Finally, it might be possible to generalize this approach to gain at least a bit of insight into the dynamics of black hole
evaporation.  The Euclidean continuation used here assumes a stationary metric, but one might artificially insert
some time dependence into $\varphi$.  From (\ref{c6}), this would add a new term to the action, with potentially
interesting implications.

\vspace{1.5ex}
\begin{flushleft}
\large\bf Acknowledgments
\end{flushleft}

This work was supported in part by Department of Energy grant
DE-FG02-91ER40674.

\vspace{1.5ex}
\begin{flushleft}
\large\bf Data availability
\end{flushleft}

Data sharing is not applicable, as no data sets were generated or analyzed in this research.

\end{document}